\documentclass[twocolumn,showpacs,preprintnumbers,amsmath,amssymb,superscriptaddress]{revtex4}

\usepackage{epsf}
\usepackage{graphicx}
\usepackage{sidecap}

\usepackage{color}

\def \beq {\begin{equation}}
\def \eeq {\end{equation}}
\begin{document}

\title{{Observation of Topological Nodal Fermion Semimetal Phase in ZrSiS}}




\author{Madhab~Neupane}
\affiliation {Department of Physics, University of Central Florida, Orlando, Florida 32816, USA}

\author{Ilya~Belopolski}
\affiliation {Joseph Henry Laboratory and Department of Physics, Princeton University, Princeton, New Jersey 08544, USA}

\author{M.~Mofazzel~Hosen}\affiliation {Department of Physics, University of Central Florida, Orlando, Florida 32816, USA}

\author{Daniel S.~Sanchez}
\affiliation {Joseph Henry Laboratory and Department of Physics, Princeton University, Princeton, New Jersey 08544, USA}

\author{Raman Sankar} \affiliation{Center for Condensed Matter Sciences, National Taiwan University, Taipei 10617, Taiwan}

\author{Maria Szlawska} \affiliation {Institute of Low Temperature and Structure Research, Polish Academy of Sciences, 50-950 Wroclaw, Poland}


\author{Su-Yang Xu}
\affiliation {Joseph Henry Laboratory and Department of Physics, Princeton University, Princeton, New Jersey 08544, USA}

\author{Klauss~Dimitri}\affiliation {Department of Physics, University of Central Florida, Orlando, Florida 32816, USA}




\author{Nagendra Dhakal}
\affiliation {Department of Physics, University of Central Florida, Orlando, Florida 32816, USA}



\author{Pablo Maldonado}
\affiliation {Department of Physics and Astronomy, Uppsala University, P. O. Box 516, S-75120 Uppsala, Sweden.}

\author{Peter M. Oppeneer}
\affiliation {Department of Physics and Astronomy, Uppsala University, P. O. Box 516, S-75120 Uppsala, Sweden.}

\author{Dariusz Kaczorowski}
\affiliation {Institute of Low Temperature and Structure Research, Polish Academy of Sciences,
50-950 Wroclaw, Poland}

\author{Fangcheng Chou} \affiliation{Center for Condensed Matter Sciences, National Taiwan University, Taipei 10617, Taiwan}

\author{M.~Zahid~Hasan}
\affiliation {Joseph Henry Laboratory and Department of Physics,
Princeton University, Princeton, New Jersey 08544, USA}

\author{Tomasz~Durakiewicz}
\affiliation {Condensed Matter and Magnet Science Group, Los Alamos National Laboratory, Los Alamos, NM 87545, USA}

\date{18 June, 2013}
\pacs{}
\begin{abstract}

{Unveiling new topological phases of matter is one of the current objectives in condensed matter physics. Recent experimental discoveries of Dirac and Weyl semimetals prompt to search for other exotic phases of matter. Here we present a systematic angle-resolved photoemission spectroscopy (ARPES) study of ZrSiS, a prime topological nodal semimetal candidate. Our wider Brillouin zone (BZ) mapping shows multiple Fermi surface pockets such as the diamond-shaped Fermi surface, ellipsoidal-shaped Fermi surface, and a small electron pocket encircling at the zone center ($\Gamma$) point, the M point and the X point of the BZ, respectively.
We experimentally establish the spinless nodal fermion semimetal phase in ZrSiS, which is supported by our first-principles calculations. Our findings evidence that the ZrSiS-type of material family is a new platform to explore exotic states of quantum matter, while these materials are expected to provide an avenue for engineering two-dimensional topological insulator systems.}




\end{abstract}
\date{\today}
\maketitle


A three-dimensional (3D) Z2 topological insulator (TI) is a crystalline solid, which is an insulator in the bulk but features spin-polarized Dirac electron states on its surface \cite{Hasan, SCZhang, Hasan_review_2, Xia, Neupane, Neupane_2, Neupane_1, TCI, Nagaosa, Young_Kane, Dai, Dai_LiFeAs}. The first 3D TI was theoretically predicted and experimentally realized in a bismuth-based compound. The discovery of the first TI tremendously accelerated research into phases of matter characterized by non-trivial topological invariants \cite{Hasan, SCZhang, Hasan_review_2, Xia}. Not only did the 3D Z2 TI itself attract great research interest, it also inspired the prediction of a range of new topological phases of matter \cite{Hasan_review_2}. The primary examples are the topological Kondo insulators, the topological 3D Dirac and Weyl semimetals, the topological crystalline insulators and the topological superconductors \cite{Hasan_review_2, Neupane_2, TCI, Neupane, Suyang_Science, Ilya_PRL, Hasan_2, Hong_Ding, TaAs_theory_1, TaAs_theory}. Each of these phases was predicted to exhibit surface states with unique properties protected by a non-trivial topological invariant. Moving ahead, new materials have been recently predicted to exhibit the topological Dirac line semimetal phase \cite{node_0, node_1, node_2, node_3, node_4}. In Weyl semimetals the bulk Fermi surface has zero dimensions, whereas in nodal line semimetals one dimensional Fermi lines in momentum space are expected.

Multiple types of nontrivial topological metallic states have been proposed for Dirac materials such as the Weyl semimetal \cite{Suyang_Science, Hong_Ding, Ilya_PRL, Hasan_2, HgCr2Se4, Vishwanath, B_B, TaAs_theory_1, TaAs_theory}, Dirac semimetal \cite{Dai, Neupane} and nodal semimetal \cite{node_0, node_1, node_2, node_3, node_4}. All of these semimetals have band-crossing points as a result of the band inversion. For Weyl and Dirac semimetals, the band crossing points which compose the Fermi surface, are located at separate  momentum space locations. Conversely, for a nodal semimetal, a closed loop is formed in the vicinity of the Fermi level, where an additional symmetry protects an extended line-like touching between the conduction and valence bands.
Since angle-resolved photoemission spectroscopy (ARPES) is  the only  momentum-resolved technique capable of isolating the surface from bulk states in topological Dirac type materials, convincing evidence of the topological nodal semimetal phase may be obtained by angle-resolved photoemission spectroscopy (ARPES), which provides an energy- and momentum-resolved probe of the electronic structure.

PbTaSe$_2$ has recently been reported as a material with Dirac line nodes \cite{Guang}. However, many bulk bands are interfering with the topological nodal line bands at the Fermi level, which prevent the manipulation and control of the topological nodal phase in this system. Moreover, the recently reported ARPES data on ZrSiS are limited because these ARPES data were obtained by using single photon energy \cite{Ast}, which prevents the experimental observation of all the possible electronic pockets at the Fermi level over a wider Brillouin zone (BZ) window. Despite many theoretical discussions of the nodal-line semimetal phase, a direct convincing experimental signature in a clean system of the nodal-line semimetal fermion phase is still lacking.

\begin{figure*}
\centering
\includegraphics[width=18.5cm]{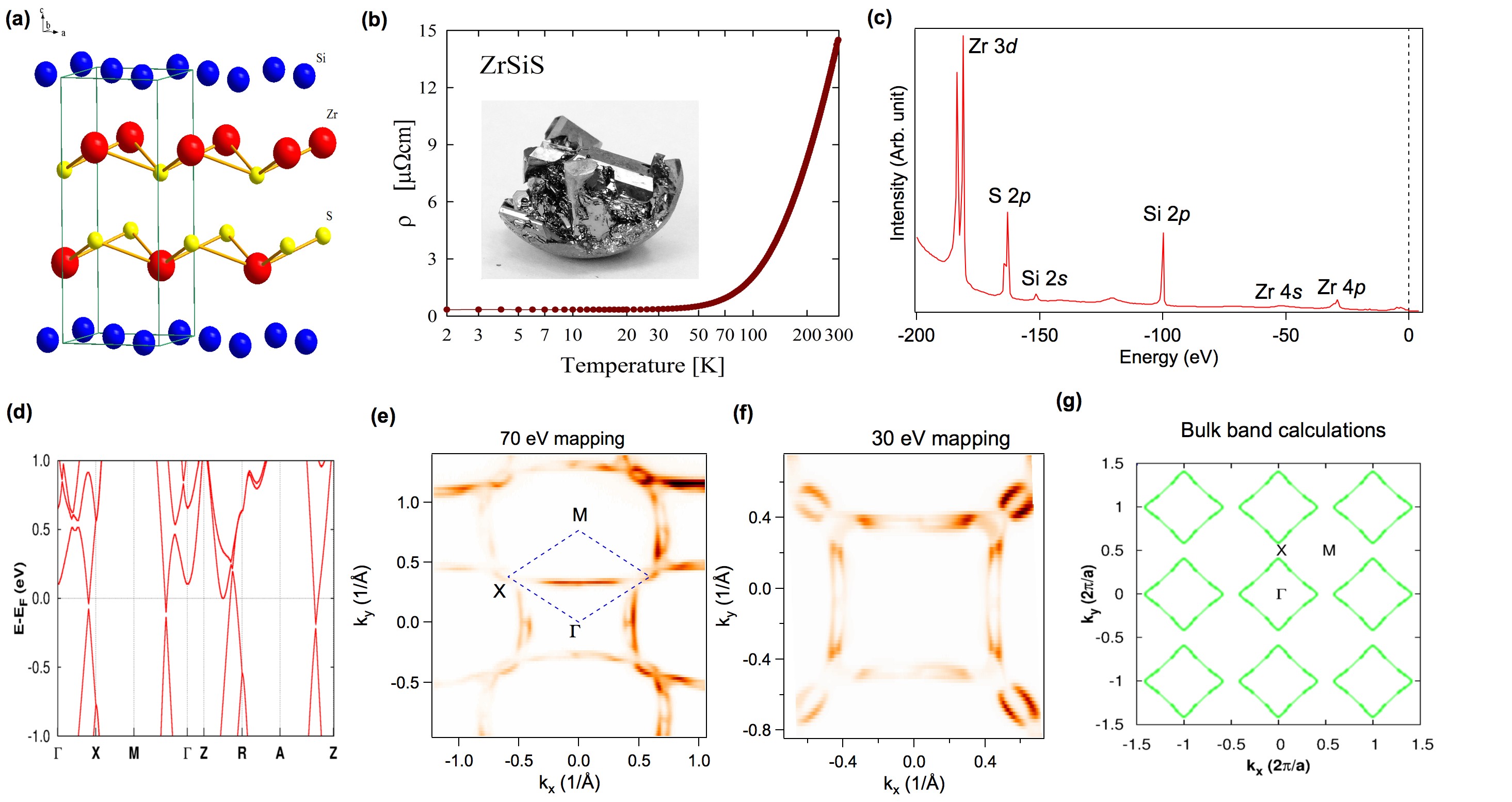}
\caption{{Crystal structure and sample characterization of ZrSiS}. 
(a) Tetragonal crystal structure of ZrSiS. The atoms form clear two-dimensional structures. Zr 
layers are separated by two neighboring S layers, where both are sandwiched between Si atoms forming 
a square net. 
(b) Temperature dependence of the electrical resistivity of single-crystalline ZrSiS measured with the electrical current flowing within the ab plane, in zero magnetic field applied along the c axis of the tetragonal unit cell. The inset shows a picture of the sample used in our measurement.
(c) Core level spectroscopic measurement of ZrSiS, which shows various core-energy levels of the constituting elements. Here the vertical dashed-line at zero energy represents the Fermi level. 
(d) Calculated band structure along various high-symmetry directions. Small-gaps are observed due to the spin-orbit coupling. (e)-(f) Measured Fermi surface maps covering multiple Brillouin zone region. Photon energies of the measurements are denoted in the plots. (g) Calculated bulk Fermi surface maps over the wider BZ regions. High-symmetry points are also marked on the plot. ARPES data were collected at the SIS-HRPES end-station at the SLS, PSI.}
\end{figure*}

In this paper, we report the experimental observation of the topological nodal fermion semimetal phase in  ZrSiS using ARPES. 
Our measurements demonstrate the topological Dirac line node phase in ZrSiS, which is further supported by our first-principles calculations. 
Our wider Brillouin zone (BZ) mapping reveals the existence of multiple pockets at the Fermi level. Specifically, we observe a diamond-shaped Fermi surface at the zone center ($\Gamma$) point and an ellipsoidal-shaped Fermi surface at  the M point and the small electron-like pockets at the X point of the BZ. Furthermore, our experimental data show that  this material possesses the linearly dispersive surface states, supported by our slab calculations. Our findings suggest ZrSiS as a new platform to explore the exotic properties of nodal fermion semimetals, a materials which favorably remains stable at air and does not contain any toxic elements in its composition. Since the ZrSiS-type of materials family consists of numerous members with more than 200 compounds \cite{Material}, our study paves the way to search for exciting quantum materials from such a large pool of material systems.

\begin{figure*}
\centering
\includegraphics[width=18.5cm]{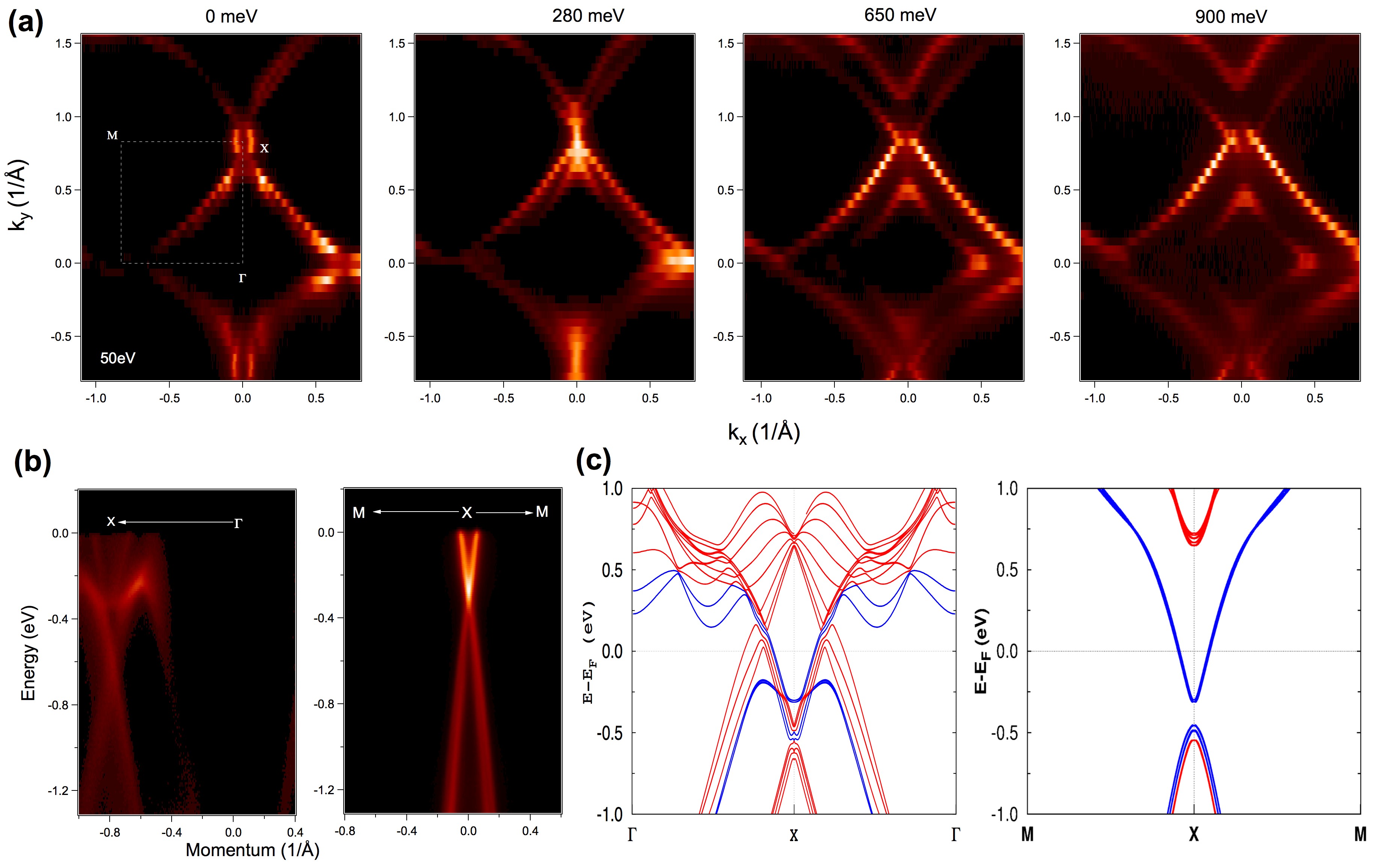}
\caption{{Electronic structure measurements of ZrSiS.} (a) Fermi surface map (with binding energy of 0 meV shown in  the top left panel) and constant energy contours. The value of the binding energy is denoted at the constant energy contour plots. (b) Dispersion map of ZrSiS along the high-symmetry directions obtained by using incident photon energy of 50 eV at a temperature of 20 K. ARPES data were collected at the SIS-HRPES end-station at the SLS, PSI. The high-symmetry directions are marked on the plots. (c) Slab calculations along high-symmetry directions. Blue and red curves represent the surface and bulk bands, respectively. Small bulk gap is observed at the X-point along the M-X-M high-symmetry direction.}
\end{figure*}

Single crystals of ZrSiS were grown by the vapor transport method as described elsewhere \cite{Material, growth}. Synchrotron-based ARPES measurements of the electronic structure  were performed at the SIS-HRPES end-station of the Swiss Light Source and ALS BL 10.0.1 with
a Scienta R4000 hemispherical electron analyzer. The energy resolution was set to be better than 20 meV  and the angular resolution was set to be better than  0.2$^{\circ}$ for the synchrotron measurements. The electronic structure calculations were carried out using the Vienna Ab-initio Simulation Package (VASP) \cite{Kress_1}, with the generalized gradient approximation (GGA) as the DFT exchange-correlation functional  \cite{Blochl_1, GGA_1}. Projector augmented-wave pseudopotentials \cite{VASP} were used with an energy cutoff of 500 eV for the plane-wave basis, which was  sufficient to converge the total energy for a given k-point sampling. In order to simulate surface effects, we used 1 $\times$ 5 $\times$ 1 supercell for the (010) surface, with a vacuum thickness larger than 19  \AA. The Brillouin zone integrations were performed on a special k-point mesh generated by 25 $\times$  25 $\times$  25 and 25 $\times$  25 $\times$  3 gamma-centered Monkhorst Pack k-point grid for the bulk and surface calculations respectively. The spin-orbit coupling was included self-consistently in the electronic structure calculations. The electronic minimization algorithm used for static total-energy calculations was a blocked Davidson algorithm.

\begin{SCfigure*}
\centering
\includegraphics[width=15.0cm]{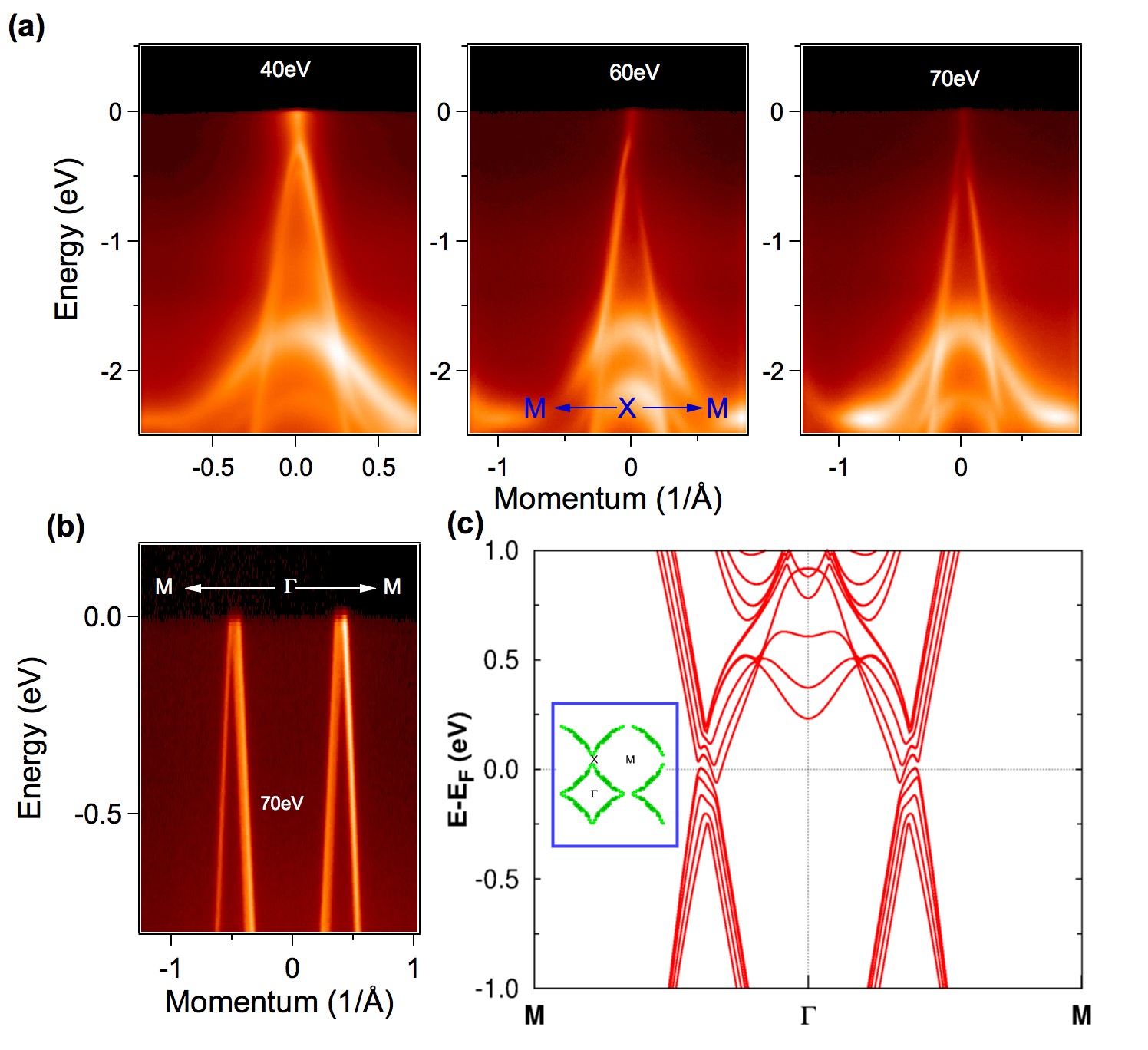}
\caption{{Dispersion maps for ZrSiS.}
(a) Photon energy dependent ARPES dispersion maps. The measured photon energies are given on the plots. The linearly dispersive states are observed, which do not show any dispersion with photon energy. (b) ARPES energy-momentum dispersion map measured with photon energy of 70 eV. (c) Calculations showing the dispersion map along the M-$\Gamma$-M high-symmetry direction. Inset shows a calculated Fermi surface formed by Dirac line nodes. Experimental data qualitatively agrees with the calculations.}
\end{SCfigure*}



We start our discussion with presenting the crystals structure of ZrSiS. It crystallizes in a PbFCl-type crystal structure with space group $P4/nmm$ \cite{Material}. In this system, Si is situated at the centre of a tetrahedron consisting of Zr atoms (see Fig. 1a). The relatively weak Zr-S bonding between two neighboring structures provides a natural cleavage plane between the adjacent ZrS layers. The crystal easily cleaves along the (001) surface, which contains Zr and S termination. Fig. 1b shows the temperature dependent resistivity of ZrSiS measured without the applied magnetic field. The zero-field dependence resistivity has a clear metallic character with residual resistivity ratio value (RRR = $\rho (300K)/\rho (2 K))$ equal to 50, that confirms the high quality of the crystal used in our measurements. The inset in Fig. 1b shows the picture of the sample used in our transport and spectroscopic measurements.


Photoemission spectroscopy (PES) also provides information about the core level states relative to the chemical potential. Figure 1c shows the core levels in the 0 - 200 eV binding energy range of the ZrSiS sample. From low to high binding energies, we observe the Zr 4$p$
($\sim$ 28 eV), Zr 4$s$ ($\sim$50 eV), Si 2$p$ ($\sim$99 eV),
Si 2s ($\sim$ 149 eV), S 2p ($\sim$ 162 eV) and Zr 3d ($\sim$ 181 eV)
states, respectively. The observation of the sharp peaks suggests that the samples used in our spectroscopic measurements are of high quality.


 
Now we discuss the electronic band structure of the ZrSiS system. 
The calculated bulk band structure along various high-symmetry directions is shown in Fig. 1d. Small-gaps are observed due to the spin-orbit coupling. The measured Fermi surface maps using photon energies of 70 eV and 30 eV are shown in Fig. 1e and Fig. 1f. These maps are obtained within the energy window of $\pm$ 5 meV near the Fermi level. Various Fermi pockets are observed in the Fermi surface map obtained by 70 eV by covering a larger area of the Brillouin zone. Specifically, a diamond shaped Fermi surface is observed around the zone center ($\Gamma$) point, and an ellipsoidal pocket is seen at around the M point. Interestingly, we also observe a small electron-like Fermi pocket around the X point of the BZ. Fig. 1g shows the calculated bulk Fermi surface map. By comparing the experimental Fermi surface maps in Figs. 1e-f with the calculated Fermi surface shown in Fig. 1g, it can be concluded that the states around the X point do not originate from the bulk band. We attribute these states to the surface electronic structure, in good agreement with  the slab calculations.


 
In order to determine the detailed electronic structure of ZrSiS, we present the results of our systematic electronic structure studies in Figs. 2 and 3. Fig. 2a shows a plot of the Fermi surface map (left) and several constant energy contours measured with various binding energies as noted in the plots, which is measured by using the photon energy of 50 eV. Experimentally we observe a diamond-shaped Fermi surface consisting of the Dirac line node (Fig. 2a left). At around 300 meV below the Fermi level, the point like state is observed at the X point, which is located around the binding energy of the Dirac point of the observed surface state (see Fig. 2a). Moving towards higher binding energy, the diamond-shaped Fermi surface having the Dirac line node becomes disconnected into inner and outer diamond-shaped structure as seen clearly in Fig. 2a (right). The dispersion maps along the high-symmetry direction at around the X point are presented in Fig. 2b. Along the $\Gamma$-X-$\Gamma$ high-symmetry direction, a Dirac like dispersion at higher binding energy ($\sim$ 600 meV) is observed, which is due to the square Si-plane and this state is protected by non-symmorphic symmetry. The observed Dirac like surface states  along the M-X-M high-symmetry direction indicates that the Dirac point is located about 300 meV below the Fermi level. 
Our experimentally observed dispersion maps qualitatively agree with the band calculations shown in Fig. 2c.
 
To reveal the nature of the states around the X point, we perform photon energy dependent ARPES measurements as shown in Fig 3a. These spectra do not disperse with the measured photon energy, which suggests that the observed states are surface states. Fig. 3b shows the dispersion map along the M-$\Gamma$-M high-symmetry direction measured with photon energy of 70 eV. These states, which come from the nodal line states, are consistent with the calculations shown in Fig. 3c. To highlight the Dirac nodal phase, we present a Fermi surface plot, shown in the inset of Fig. 3c, formed by the nodal lines. We ignore the small spin-orbit coupling gap for the calculation (it is estimated to be less than 10 meV which is below the resolution of typical ARPES measurements) and thus the calculated nodal Fermions are spinless.  We further note that we do not observe the SOC gap in experimental data, suggesting a negligible size of the SOC gap.
To highlight the Dirac nodal phase, we illustrate the nodal lines for a specific plane of the BZ (k$_z$ = 0) in the inset of Fig. 3c. Here, the Dirac line nodes connect the Dirac points at the Fermi level.
Furthermore, we observe that monolayer systems of ZrSiS type materials are interesting because these might show edge state protected by non-symmorphic space group symmetry \cite{Young_Kane}.

 In conclusion, we have performed systematic ARPES measurements on the  ZrSiS system covering a large area of the Brillouin zone. We reveal the existence of multiple Fermi pockets and identify their origin, which is complemented by our Ab-initio calculations. Importantly, we  reveal  the existence of the topological spinless nodal semimetal fermion phase in this system, which emerges because of a negligible SOC gap. Our study establishes the ZrSiS family of materials as a new platform to study the Dirac line node semimetal phases, with the possibility of realizing novel exciting quantum phases utilizing the large pool of material systems available in this family.

\bigskip
\bigskip
\bigskip
\hspace{0.5cm}
\textbf{Acknowledgements}
\newline

M.N. is supported by the start-up fund from the University of Central Florida.
T.D. is supported by NSF IR/D program. 
I. B. acknowledges the support of the NSF GRFP.
D.K. was supported by the National Science Centre (Poland) under research grant 2015/18/A/ST3/00057.
Work at Princeton University is supported by the Emergent Phenomena in Quantum Systems Initiative of the Gordon and Betty Moore Foundation under Grant No. GBMF4547 (M.Z.H.) and by the National Science Foundation, Division of Materials Research, under Grants No. NSF-DMR-1507585 and No. NSF-DMR-1006492. P.M. and P.M.O. acknowledge support from the Swedish Research Council (VR), the K. and A. Wallenberg Foundation and the Swedish National Infrastructure for computing (SNIC). We thank Plumb Nicholas Clark for beamline assistance at the SLS, PSI. We also thank Sung-Kwan Mo for beamline assistance at the LBNL.








\*Correspondence and requests for materials should be addressed to M.N. (Email: Madhab.Neupane@ucf.edu).

\end{document}